\documentclass[%
reprint,
superscriptaddress,
 amsmath,amssymb,
 aps, pre,
]{revtex4-2}

\usepackage{graphicx}
\usepackage{afterpage}
\usepackage{float}
\usepackage{dcolumn}
\usepackage{bm}
\usepackage[colorlinks=true,linkcolor=blue,urlcolor=blue,citecolor=blue,anchorcolor=blue]{hyperref}
\usepackage{adjustbox}

\newcommand{\modified}[1]{{#1}}

\begin{document}

\title{Higher-order dissimilarity measures for hypergraph comparison}

\author{Cosimo Agostinelli}
\affiliation{Aix-Marseille Univ, Université de Toulon, CNRS, CPT,
Turing Center for Living Systems, 13009 Marseille, France}
\author{Marco Mancastroppa}
\affiliation{Aix-Marseille Univ, Université de Toulon, CNRS, CPT,
Turing Center for Living Systems, 13009 Marseille, France}
\author{Alain Barrat}
\email{alain.barrat@cpt.univ-mrs.fr}
\affiliation{Aix-Marseille Univ, Université de Toulon, CNRS, CPT,
Turing Center for Living Systems, 13009 Marseille, France}

\begin{abstract}
{In recent years, networks with higher-order interactions have emerged as a powerful tool to model complex systems.
Comparing these higher-order systems remains however a challenge. Traditional similarity measures designed for pairwise networks fail indeed to capture salient features of hypergraphs, hence 
potentially neglecting important information.
To address this issue, here we introduce two novel measures, Hyper NetSimile and Hyperedge Portrait Divergence, specifically designed for comparing hypergraphs. 
These measures take explicitly into account the properties of 
multi-node interactions, using complementary approaches. They 
are defined for any arbitrary pair of hypergraphs, of potentially different sizes,
thus being widely applicable.
We illustrate the effectiveness of these metrics through clustering \modified{tasks} on synthetic and empirical higher-order networks, showing their ability to correctly group hypergraphs generated by different models and to distinguish real-world systems belonging to different contexts.
Our results highlight the advantages of using higher-order dissimilarity measures over traditional pairwise representations in capturing the full structural complexity of the systems considered.}

\end{abstract}

\maketitle

\section{Introduction} 
Many systems in diverse domains can be effectively described using networks, in which fundamental elements are represented as nodes and their interactions as links connecting these nodes \cite{barabasi2012,newman_book,barrat2008dynamical}. This modeling framework has been highly influential in analyzing and understanding a variety of complex phenomena, ranging from transportation networks to biological interactions and epidemics \cite{vespignani2012,barrat2008dynamical}.
Within this framework, different systems are represented through a common schematic approach, allowing in particular to quantify the similarities between these systems by 
measuring the discrepancies between their representations. 
To this aim, several tools and techniques for defining network similarity measures have been proposed. These measures generally leverage either structural properties and their statistics, such as centrality distributions and graphlet statistics \cite{yaveroglu2014, Berlingerio2012}, 
statistics of paths and distances between nodes in the network \cite{schieber2017quantification,Bagrow2019}, or spectral properties of the network \cite{shimada2016}. 
Such similarity measures \cite{McCabe2021, tantardini2019, willis2020} have a broad range of practical applications and have been employed in several disciplines, for instance, in the classification of biological structures \cite{Sharan2006} and in the analysis of the evolution of social systems \cite{masuda_holme2019,gelardi2019detecting}.

However, recently it has been shown that this framework, despite its utility, presents some intrinsic limitations: by definition, networks can only describe systems where elements interact in pairs and, as such, they fail to capture the multi-body interactions that drive many real-world phenomena, from chemical reactions \cite{jost2019} to social and ecological systems \cite{benson2016, grilli2017}.
To overcome this issue and take into account group interactions, more general mathematical frameworks can be used, such as hypergraphs and simplicial complexes \cite{bick2023,battiston2020}. These higher-order representations extend the descriptive power of pairwise graph theory, enabling the investigation of multi-node interactions and revealing behaviors that remain hidden when reduced to pairwise approximations
\cite{iacopini2019,battiston2021physics}. 
These mathematical structures have emerged as a new paradigm for modeling complex systems thanks to their descriptive power and their rich phenomenology \cite{bick2023,battiston2020,iacopini2019, skardal2020,battiston2021physics,ferrazdearruda2023}. 

In this context, a natural and pressing challenge arises: how can we compare these higher-order systems? Is it useful to consider intrinsically higher-order features in the comparison, or is it enough to use tools defined on pairwise networks?
A simplistic approach to this challenge would indeed be to convert the interactions among groups of nodes (called hyperedges) into the corresponding sets of pairwise links (which become cliques in the resulting network). However, this projection eliminates the higher-order structure of the system, thus potentially leading to an important loss of structural information and a lower ability to distinguish between similar systems.
For example, two different higher-order networks with the same projection would be classified as identical by this procedure. Using tools defined for pairwise networks is thus not sufficient, and 
measures of similarity taking explicitly into account the higher-order properties of hypergraphs are needed.
Very few attempts have however been made to define such similarity measures \cite{surana2023, Feng2024}, 
and principled and effective methods are still lacking.

Here we devise two measures to quantify the similarities between hypergraphs, that we call
respectively Hyper NetSimile and Hyperedge Portrait Divergence. These
methods are inspired by concepts initially proposed for comparing pairwise networks \cite{Berlingerio2012,Bagrow2019}, but take
explicitly into account multi-node interactions, and can be used to compare any pair of hypergraphs, even of different sizes, as they do not rely on node correspondences between the structures being compared.
The measures we propose feature a high discriminatory power, being able to cluster correctly a set of synthetic hypergraphs generated by different models
(i.e., in groups corresponding each to one of the models), as larger similarities
are obtained between hypergraphs generated by the same underlying generation mechanisms.  
Empirical hypergraphs can also be clustered effectively according to the type 
of real-world system they represent.
Moreover, the metrics we put forward make it possible to distinguish between 
different types of randomization procedures applied to a given hypergraph.
To highlight the interest and need of using similarity measures leveraging 
higher-order properties, we systematically consider the 
results obtained by neglecting the group interactions and representing the systems
by networks (projecting the hyperedges onto network cliques). We
find that taking into account explicitly higher-order interactions generally leads to better performances than
using only pairwise representations and measures.

\section{Results} 

We consider a hypergraph $\mathcal{H} = (\mathcal{V},\mathcal{E})$, where $\mathcal{V}$
is the set of its $|\mathcal{V}|=N$ nodes and $\mathcal{E} \subseteq \{e : e \subseteq \mathcal{V}\}$ is the set of its $|\mathcal{E}|=E$ hyperedges, that represent the interactions among groups of nodes. We refer to the cardinality of a hyperedge $|e|$ as the size of that interaction, and the maximum hyperedge size in $\mathcal{H}$ is denoted by $M$. 
Throughout this work we will consider undirected and unweighted hypergraphs for simplicity.
We moreover call \textit{pairwise projection} of $\mathcal{H}$ the network $\mathcal{G(H)}$ obtained by replacing all the hyperedges in $\mathcal{H}$ with the corresponding set of pairwise links, each hyperedge giving rise to a clique. For example, a hyperedge $(1,2,3)$ in $\mathcal{H}$ is replaced by the
three links $(1,2),\ (2,3),\ (1,3)$ in $\mathcal{G(H)}$. For simplicity, here we neglect multiple edges in $\mathcal{G(H)}$ that may arise due to overlap between hyperedges in $\mathcal{H}$.
\modified{For completeness however, we also consider in the Supplementary Information (SI) a \textit{weighted projection} of the hypergraphs, where each edge $(i,j)$ is assigned a weight corresponding to the number of different hyperedges in $\mathcal{H}$ involving both $i$ and $j$.
}

Given two hypergraphs $\mathcal{H}_1, \mathcal{H}_2$ our goal is to define a measure $d(\mathcal{H}_1, \mathcal{H}_2)$ that quantifies the dissimilarity between $\mathcal{H}_1, \mathcal{H}_2$. The concept of dissimilarity (or similarity) is however arbitrary and depends on the features of the systems in which one is interested. 
In particular, as for the case of dissimilarities between networks, we consider structural aspects of the hypergraphs, and we define two dissimilarity measures based on two different characterization approaches. 
In the first metric we propose, which we call \textit{Hyper NetSimile},
we consider a set of relevant node features 
such as the hyperdegrees and the sizes of hyperedges in which the nodes are involved. It has a node-centric perspective, so that every feature is measured either at the level of single nodes or at the level of their neighborhood, 
\modified{up to distance two}.
In the second metric, \textit{Hyperedge Portrait Divergence}, 
we focus on the statistics of paths and distances within the hypergraph, thus accounting also for the diffusive properties of the system. Moreover, in this case we shift to a hyperedge-centric perspective, considering the hyperedges as the building blocks of a higher-order network, and defining a measure that relies on hyperedge-based paths.

\subsection{Higher-order dissimilarity measures}

\subsubsection{Hyper NetSimile} 
The first measure we introduce is a generalization of NetSimile (NS), a metric originally proposed for comparing pairwise networks \cite{Berlingerio2012}. The idea is to associate a feature vector to each hypergraph $\mathcal{H}$, and then to take the distance between vectors as an indicator of dissimilarity between the related hypergraphs. 
To build such vector, we consider the distributions of the following quantities over all the nodes $i$ in $\mathcal{H}$: 
\begin{enumerate}
    \item number of $i$'s neighbors;
    \item $i$'s hyperdegree, i.e., the number of hyperedges $i$ belongs to;  
    \item average size of the hyperedges containing $i$;
    \item standard deviation of the size of hyperedges containing $i$;
    \item average number of neighbors of $i$'s neighbors;
    \item average hyperdegree of $i$'s neighbors;
    \item number of neighbors of $i$'s ego-net, i.e., number of nodes that are two steps away from $i$
    on $\mathcal{H}$.
\end{enumerate}
\modified{For all these distributions, we compute three statistical indicators -- mean, median, and standard deviation -- and concatenate them to obtain a signature vector $\bm{v}$ for each hypergraph, here of length $V=21$. 
These vectors contain information about the most important local properties of the systems as they capture the structure of each node's neighborhood, up to a distance two}.
Note that the features \modified{1, 5, and 7} give the same values when measured on a hypergraph or on the corresponding projected network, i.e., the network obtained by replacing the higher-order interactions with pairwise cliques.
\modified{The other features are instead purely higher-order and become either trivial or redundant if applied to the projected network.}
Finally, following the original pairwise formulation \cite{Berlingerio2012}, we choose the Canberra distance to compute the difference between the signature vectors of the two hypergraphs we want to compare.
\modified{This distance has the advantage of being quite sensitive to small changes and normalizable, a necessary property for relating distances among multiple couples of elements with one another \cite{Berlingerio2012}.}
The Hyper NetSimile (HNS) dissimilarity between $\mathcal{H}_1$ and $\mathcal{H}_2$ (represented respectively through the signature vectors $\bm{v_1}$ and $\bm{v_2}$) is thus given by:
\begin{equation}
    HNS(\mathcal{H}_1, \mathcal{H}_2) = d_{canberra}(\bm{v_1}, \bm{v_2}) = \frac{1}{V} \sum_{j=1}^{V} \frac{|v_1^j - v_2^j|}{|v_1^j|+|v_2^j|},
    \label{eq. HNS}
\end{equation}
where we have normalized the distance by the length $V$ of the vectors, so that $HNS(\mathcal{H}_1, \mathcal{H}_2) \in [0,1]$ $\forall \, \mathcal{H}_1, \mathcal{H}_2$.

\modified{Note that the original formulation of NS includes also skewness and kurtosis among the statistical indicators used to build the feature vectors.
We decided not to include them in the definition of HNS because they actually worsen the performances of the metric (see SI). This can be intuitively understood for the skewness, that measures the degree of asymmetry of a probability distribution and can take both positive and negative values. 
The sum in the Canberra distance used in Eq. \eqref{eq. HNS} returns the maximum possible distance for the entry $j$ if $v_1^j$ and $v_2^j$ have opposite sign, regardless of their actual difference,
meaning that this indicator can produce a large dissimilarity between two distributions even if they are not particularly different from each other.
In the following we will also consider a version of NS without skewness and kurtosis, in order to have a fairer comparison between HNS and NS.
In the SI we present some results obtained with HNS and NS when skewness and kurtosis are included in their definition, showing that the performance of both decreases slightly.
}

\modified{Similarly, when listing the features to consider in the hypergraphs, we have not
considered the clustering coefficient.
The motivation is that, although there are principled generalization of this coefficient for hypergraphs \cite{Zhou2011}, they are computationally expensive.
Moreover, we have checked (see SI) that by considering features based on the clustering coefficient, the performances of HNS get worse instead of improving.
}

\subsubsection{Hyperedge Portrait Divergence}
The second measure we introduce is a generalization of the Portrait Divergence (PD), a measure proposed for networks comparison \cite{Bagrow2008,Bagrow2019}. Given a network $\mathcal{G}$, its \textit{portrait} is defined as the matrix $B$ whose element $B_{l,k}$ is the number of nodes of $\mathcal{G}$ having $k$ nodes at distance $l$ \cite{Bagrow2008}. This matrix contains information about several properties of the system, such as the total number of nodes and degree distribution, and overall encodes information based on paths of all lengths on the network \cite{Bagrow2019}. 
It is \modified{then converted into} the probability $P(l,k)$ of randomly choosing a pair of nodes that are distant $l$ from one another, so that one of them has $k$ nodes at distance $l$: the Portrait Divergence (PD) between two networks is then defined as the Jensen-Shannon divergence between their corresponding $P(l,k)$ \cite{Bagrow2019}.

When considering higher-order systems, it is important to take into account the different sizes of interaction and, to this aim, we propose to consider the hyperedges as the basic elements of a hypergraph instead of the nodes and to use a definition of a portrait that takes
hyperedge sizes into account. 
We thus define the \textit{Hyperedge Portrait} $\Gamma$ of $\mathcal{H}$ as a tensor with four indices, whose component $\Gamma_{m,n,l,k}$ is given by 
the number of hyperedges of size $m$ having $k$ hyperedges of size $n$ at distance $l$. 
We consider that two distinct hyperedges are at distance $1$ if they share at least one node, and in this case they are said to be adjacent; respecting additivity, we compute the distance between hyperedges as the length of the shortest hyperedge-path connecting them, while moving through adjacent hyperedges. This is equivalent to measure the distance between hyperedges in the bipartite representation of the hypergraph.
Akin to the network portrait $B$, also $\Gamma$ encodes relevant characteristics of the system it represents. For example, we can recover the number of hyperedges of size $s$, denoted by $E_s$:
\begin{equation} 
    \Gamma_{s,n,0,k} = \delta_{s,n}\delta_{k,1}E_{s} + (1-\delta_{s,n})\delta_{k,0}E_{s},
    \label{Gamma_prop}
\end{equation}
where $\delta_{ij}$ is the Kronecker delta.
To build a dissimilarity metric from this tensor, we simply normalize $\Gamma$ and interpret it as a probability distribution $P(m,n,l,k)=\Gamma_{m,n,l,k}/\sum_{m,n,l,k}\Gamma_{m,n,l,k}$: we compute the dissimilarity between two hypergraphs as the Jensen-Shannon divergence between their respective $P(m,n,l,k)$. 
This measure of statistical difference between distributions presents many desirable aspects, such as the normalization in the interval $[0,1]$. 
The Hyperedge Portrait Divergence (HPD) between two hypergraphs $\mathcal{H}_1,\ \mathcal{H}_2$ is thus given by:
\begin{equation}
    HPD(\mathcal{H}_1,\mathcal{H}_2) = JS \big[ P_1(m,n,l,k),\ P_2(m,n,l,k) \big] \ .
    \label{eq. HPD}
\end{equation}

The two measures we propose, HNS and HPD, are based on two different approaches (set of features vs. properties of paths at all scales, node-centric perspective vs. hyperedge-centric perspective) that are complementary to each other. Therefore, these two metrics may be used together to assess the dissimilarity between higher-order networks from a twofold point of view.
Moreover, they both display four desirable properties: \textit{(i)} ease of interpretation; \textit{(ii)} possibility to compare any arbitrary couple of hypergraphs, even with a different number of nodes and \modified{hyperedges}; \textit{(iii)} normalization in the interval $[0,1]$; \textit{(iv)} invariance under relabeling of the nodes. 
The ease of interpretation of the quantities on which each measure is based (features or paths)
makes it also possible to check if the measures behave as expected in some baseline cases and control scenarios, where we have a prior knowledge and intuition about how similar to each other two systems are.

\subsection{The need for higher-order measures}
To illustrate the need for higher-order similarity measures, we first use our metrics in contexts where pairwise methods would fail by construction. 
To this aim, we consider examples of pairs of 
hypergraphs $\mathcal{H}_1$ and $\mathcal{H}_2$ sharing the same pairwise projection. The fact that $\mathcal{G(H}_1)=\mathcal{G(H}_2)$ implies indeed
that any pairwise metric would detect a dissimilarity equal to 0, since 
the differences between $\mathcal{H}_1$ and $\mathcal{H}_2$ are purely higher-order. 

As a first example of such cases, we consider a reference hypergraph $\mathcal{H}$ and randomly project a fraction $f$ of its hyperedges to the corresponding pairwise interactions, obtaining $\mathcal{H}_{null}(f)$. This hypergraph plays the role of a null model against which we can test the sensitivity of HNS and HPD, since by definition its projected graph is equal to the one of 
$\mathcal{H}$: $\mathcal{G(H})=\mathcal{G}(\mathcal{H}_{null}(f))$. 

\begin{figure*}[ht!]
    \includegraphics[width=0.97\textwidth]{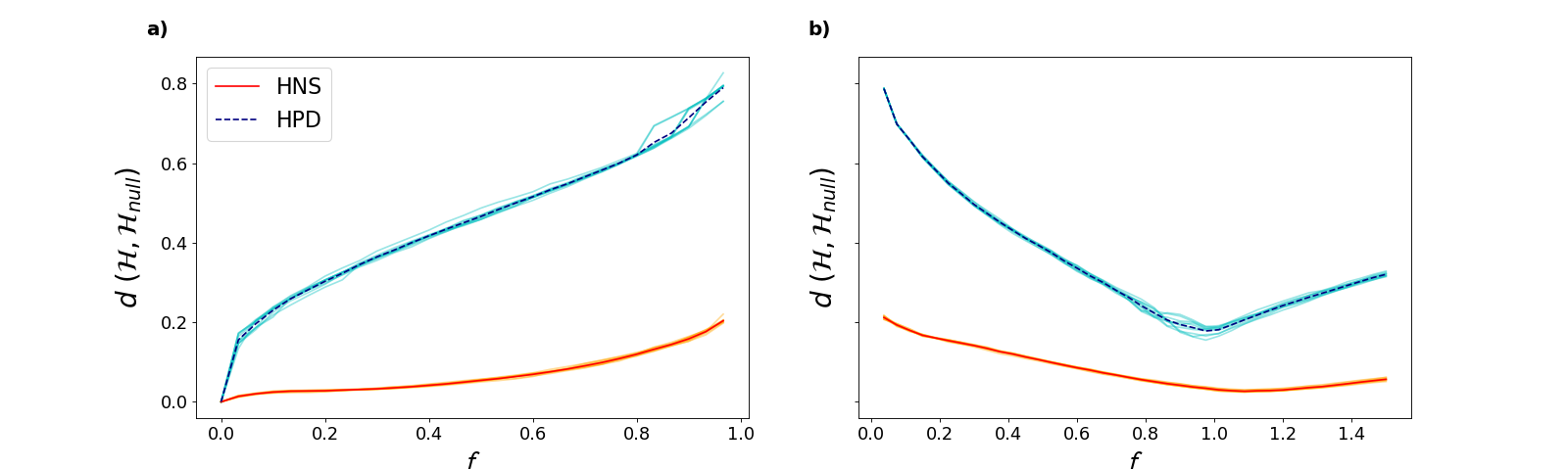}
    \caption{\label{fig_prj_null_models}\textbf{Dissimilarity between a baseline hypergraph $\mathcal{H}$ and its projection-preserving null models $\mathcal{H}_{null}(f)$}. The dissimilarity is represented as a function of the fraction $f$ of projected/promoted hyperedges. (a) $\mathcal{H}_{null}(f)$ is obtained by randomly projecting a fraction $f$ of hyperedges of $\mathcal{H}$ to pairwise edges. (b) $\mathcal{H}_{null}(f)$ is obtained by randomly promoting to hyperedges a fraction $f$ of the cliques of $\mathcal{G(H)}$. The red solid and blue dashed curves (for HNS and HPD respectively) are averaged over 10 realizations of $\mathcal{H}_{null}(f)$. Single realizations are drawn in orange and light blue. The reference hypergraph $\mathcal{H}$ is built from a data set of face-to-face interactions, collected in a hospital by the SocioPatterns collaboration \cite{sociopatterns, genois2018, Vanhems2013}.}
\end{figure*}

To illustrate the procedure, we
consider an empirical hypergraph $\mathcal{H}$ built from a data set of face-to-face interactions collected within a hospital by the SocioPatterns collaboration \cite{sociopatterns, genois2018, Vanhems2013} (see Methods for more details on data collection and preprocessing).
The dissimilarity between $\mathcal{H}$ and its partially projected version
$\mathcal{H}_{null}(f)$ is shown in Fig. \ref{fig_prj_null_models}a as a function of $f$. 
Although the underlying dyadic network $\mathcal{G}(\mathcal{H}_{null}(f))$ does not depend on $f$, the dissimilarity between $\mathcal{H}$ and $\mathcal{H}_{null}$ grows monotonically with $f$ for both HNS and HPD, indicating that structural differences can be detected if the higher-order interactions are taken into account, even if paiwise measures would be unable to do so. 
HPD takes greater values than HNS, increasing sharply for small $f$ values.
This is due to the definition of the hyperedge portrait: on the one hand, it explicitly relies on the sizes of hyperedges (which are strongly modified by the projection); on
the other hand, it encodes statistics of path lengths, which are impacted by the projection of
even few hyperedges (as each clique yields a set of paths of length $1$ between hyperedges of size $2$). These two characteristics make this measure very sensitive to the partial projection procedure.

The second example we consider is complementary to the former.
We start from the pairwise projection $\mathcal{G(H)}$ of a given hypergraph $\mathcal{H}$
and randomly promote a number $f E_s$ of cliques of size $s$ of $\mathcal{G(H)}$ to hyperedges
(recall that $E_s$ is the number of hyperedges of size $s$ in $\mathcal{H}$). We perform this promotion procedure for every size of interaction $s$ represented in $\mathcal{H}$, obtaining the new hypergraph $\mathcal{H}_{null}(f)$. Note that when $f=1$ we obtain the same number of $s$-hyperedges in $\mathcal{H}$ and $\mathcal{H}_{null}$ for every $s$. However, the nodes involved in those interactions will
in general not be the same in $\mathcal{H}$ and $\mathcal{H}_{null}(f=1)$.
Figure \ref{fig_prj_null_models}b shows the dissimilarity between the original hypergraph and this second null model, as a function of the fraction $f$ of promoted cliques (the starting empirical hypergraph considered is the same as in Fig. \ref{fig_prj_null_models}a). 
It is interesting to notice that the dissimilarity is minimum around $f=1$, meaning that the closest instance of the null model to $\mathcal{H}$ is the one preserving the number and sizes of hyperedges, as could be expected. Again, this effect is more pronounced for HPD, given the sensitivity of the hyperedge portrait to the statistics of hyperedge sizes.


These two examples illustrate how the metrics we propose are both able to distinguish between hypergraphs that differ only in purely higher-order properties, thanks to the fact that these metrics are built using higher-order information and not only pairwise statistics.


\subsection{\label{section_Hmodels}Clustering of hypergraph model instances}

A way to investigate the effectiveness of similarity measures is to test them on a clustering task. To this aim, we evaluate the ability of the proposed metrics to distinguish synthetic hypergraphs generated through different models. We consider three generative models of hypergraphs that generalize standard network models \cite{Bollobas2001, newman_book, WS}. Given a set of $N$ nodes, they are defined as follows (see Methods for more details):
\begin{itemize}
    \item \textit{Erd\H{o}s-Rényi (ER)}: the hyperedges connecting $s$ nodes are randomly created with a certain probability $p_s$ that depends only on the size of the interaction. This model generates a random structure, with all nodes behaving similarly (e.g., similar hyperdegrees and neighbourhood properties).
    \item \textit{Configuration Model (CM)}: for each size $s$, the nodes are selected to participate in hyperedges to reproduce a specific $s$-degree sequence. We choose a power-law-like degree distribution for every $s$ as the characteristic feature of this model, which hence presents heterogeneity in nodes properties.
    \item \textit{Watts-Strogatz (WS)}: the starting structure is a ring lattice, where adjacent nodes are connected to each other up to a fixed distance, that depends on the size $s$ of the interactions. The hyperedges are then randomly rewired according to a certain probability $p_{rew}$. This mechanism generates a small-world effect, reducing the average distance between nodes with respect to the initial ring, while retaining large clustering values.
\end{itemize}
\modified{We generate $100$ instances of each model, using sizes of interaction up to $7$ for the ER model,
up to $3$ for the CM, and up to $5$ for the WS hypergraphs.
The number of nodes $N$ is chosen uniformly at random in the interval $[200,300]$ for each instance. Similarly, the other model parameters are randomly selected within a fixed range for every realization of the models (see Methods for further details).
}
We then compute the dissimilarity between all the pairs of the $300$ generated hypergraphs, using 
the metrics HNS and HPD. We also project each instance on its pairwise version and compute
the dissimilarities NS and PD between all the pairs of resulting graphs.

\begin{figure*}[t!]
    \includegraphics[width=\textwidth]{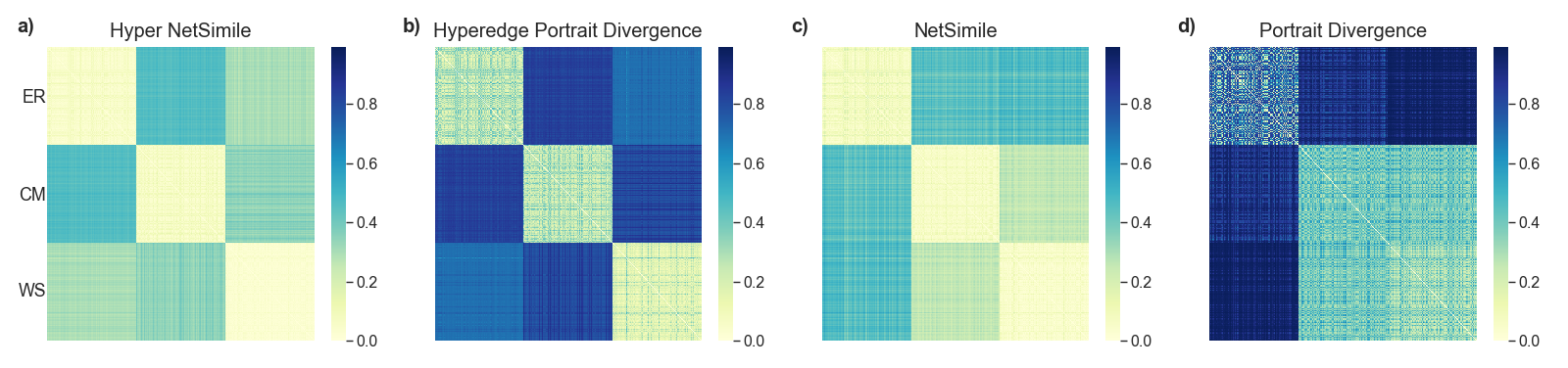}
    \caption{\label{fig_sim_matrices_models} \textbf{Dissimilarity matrices of hypergraphs models.} The matrices are computed with higher-order (a-b) and pairwise (c-d) metrics. The elements are sorted by model, so that the three blocks around the diagonal correspond to dissimilarities between instances of 
    the same model (respectively
    the Erd\H{o}s-Rényi (ER), the Configuration (CM), and the Watts-Strogatz (WS) models). 
    For each model we sample $100$ realizations: each realization has a number of nodes chosen in the interval $[200,300]$ and the maximum hyperedge size is $M=7$ for the ER model, $M=3$ for the CM model, and $M=5$ for the WS model. The other models' parameters are selected uniformly at random for each realization: the average $s$-degree of nodes $\langle k_s \rangle = \bar{k} \in [1.5,2]$ $\forall s$ for the ER model; the exponent of the degree distributions $\gamma\in [2,2.05]$ for the CM model; the rewiring probability $p_{rew}\in [0.15,0.2]$ for the WS model.}
\end{figure*}

This procedure yields four $300\times 300$ symmetric matrices, one for each metric (higher-order and pairwise), depicted in Fig. \ref{fig_sim_matrices_models}.
\modified{A visual inspection of these matrices shows that three diagonal blocks corresponding to the generative models clearly emerge for HNS, HPD, and NS, while PD is not able to distinguish the CM from the WS hypergraphs.}
To quantitatively compare the performances of the different measures, we apply a clustering algorithm on the dissimilarity matrices and evaluate how the groups given by the algorithm match the generative models. This can be done by considering the Rand Index, $RI$ \cite{RI}, and the Dunn Index, $DI$ \cite{DI}:
the Rand Index quantifies the similarity between the clusters found by the algorithm and the
ground truth groups (here, corresponding to the three models), with $RI=1$ for a perfect
correspondence; the Dunn Index is a measure of the quality of the result of the clustering algorithm that instead does not depend on the ground truth, 
as it quantifies the degree of separation between groups (see Methods for the precise definitions).
We consider here two versions of agglomerative clustering algorithm (fixing the numbers of clusters to $3$ in order to match the number of models). 
The first one computes the distance between two clusters $c_1,\ c_2$ as the minimum distance between their elements, that is, $d(c_1,c_2)=\min_{i\in c_1,\ j\in c_2} d(i,j)$, while the second one considers the average distance $d(c_1,c_2)=\langle d(i,j) \rangle _{i\in c_1,\ j\in c_2}$.
Table \ref{tab:RI_DI_models} gives the resulting values of the Rand and Dunn indices for each dissimilarity measure and for each clustering algorithm.

The three metrics NS, HNS and HPD perfectly recover the grouping corresponding to the three generative models ($RI=1$).
Using PD as dissimilarity measure makes it instead impossible to correctly recover the correct
separation of the model instances in three groups, as anticipated above.
\modified{Furthermore, HNS and HPD display larger values of the $DI$ with respect to NS, suggesting that higher-order measures can better separate the clusters from each other.
}

In the SI we consider the ability of the measures to discriminate instances of hypergraphs obtained 
within the same model class and with different model parameters. We focus on one model (the ER) and repeat the previous analysis while varying the maximum size of hyperedges $M$. 
\modified{We find that only HNS and HPD can effectively separate the hypergraphs according to the value of $M$, while the other metrics fail, as they are more impacted by the underlying pairwise structure which remains random.
We also consider in the SI 
the weighted projection of the hypergraph models and suitable extensions of the NS and PD metrics for weighted graphs \cite{Berlingerio2012,Bagrow2019}. In this case the performance of NS and PD improves, as one could expect since this type of projection retains more information about the original higher-order structure than the mere pairwise approximation.
Nevertheless, the best classification is still given by HNS and HPD, underscoring 
the importance of taking into account higher-order interactions.
}

\begin{table}[t!]
    \begin{ruledtabular}
        \begin{tabular}{ccccc|cccc}
             &\multicolumn{4}{c}{minimum} \vline &\multicolumn{4}{c}{average} \\ \hline
            &HNS &HPD &NS &PD &HNS &HPD &NS &PD\\ \hline
             RI &1.00 &1.00 &1.00 &0.78 &1.00 &1.00 &1.00 & 0.74 \\
             DI &1.48 &1.09 &0.78 &0.32 &1.48 &1.09 &0.78 &0.17 \\
        \end{tabular}
    \end{ruledtabular}
    \caption{Rand Index ($RI$) and Dunn Index ($DI$) for the clustering of model-generated hypergraphs (see Fig. \ref{fig_sim_matrices_models}). The clusters can depend both on the dissimilarity metric (HNS, HPD, NS, PD) considered and the algorithm (``minimum" or ``average").}
    \label{tab:RI_DI_models}
\end{table}

\subsection{Clustering of randomized hypergraphs}

\begin{figure*}[hbt!]
    \includegraphics[width=\textwidth]{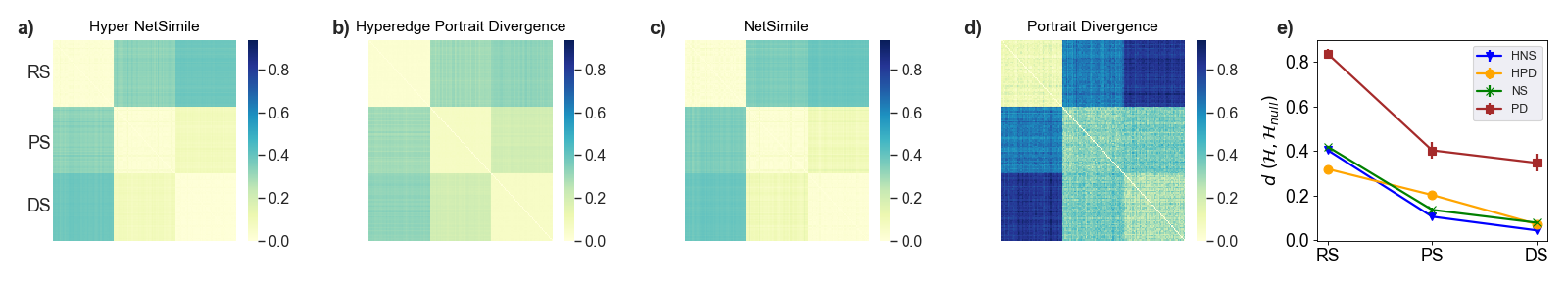}
    \caption{\label{fig_sim_matrices_resh_LH10} \textbf{Dissimilarity matrices between all pairs
    of randomized hypergraphs.} We sample 50 realizations of each method, namely Random (RS), Proportional (PS), and Degree-preserving (DS) Shuffling, and compute the dissimilarity between each pair of realizations, both with higher-order (a-b) and pairwise (c-d) metrics. Panel (e) displays the average distance between the original hypergraph and the realizations of the three randomization methods, as computed by the various metrics. The error bars represent the standard deviation \modified{(not always visible, if smaller than the marker size)}. The original hypergraph is built from face-to-face interactions data collected in a hospital by the SocioPatterns collaboration.}
\end{figure*}

We further test the sensitivity of the proposed metrics by comparing three types of randomization methods applied to a reference hypergraph $\mathcal{H}$.
As in Fig. \ref{fig_prj_null_models}, we consider the empirical hypergraph built from face-to-face interactions data, recorded in a hospital by the SocioPatterns collaboration \cite{genois2018, sociopatterns, Vanhems2013}.
We then randomize the hyperedges of this hypergraph $\mathcal{H}$ according to three different methods (see Methods):
\begin{itemize}
    \item \textit{Random Shuffling (RS)}: this randomization keeps only the number and size of the hyperedges of the original system, while the nodes belonging to each hyperedge are chosen uniformly at random, destroying node heterogeneity and the hypergraph structure. 
    \item \textit{Proportional Shuffling (PS)}: this method is similar to the RS, but the nodes in each hyperedge are selected with probability proportional to their hyperdegree in the original hypergraph. Hence, the hyperdegrees in the original hypergraph and in its randomized version will be approximately the same for each node. Note that this notion of hyperdegree does not take into account the sizes of the hyperedges in which a node takes part.
    \item \textit{Degree-preserving Shuffling (DS)}: this procedure shuffles the hyperedges while keeping fixed the original hyperdegree of every node at every order of interaction. Thus, it leaves the hyperdegree statistics unchanged at every order of interaction, while still affecting, as the other methods, the meso- and large-scale structure of the system (e.g., destroying communities and hierarchical structures \cite{mancastroppa2023hyper}).
\end{itemize}
For each randomization method we sample 50 realizations of $\mathcal{H}_{null}$ and perform the same analysis presented in the previous Section, i.e., we compute the dissimilarity between all pairs of randomized instances, using both HNS and HPD. Each realization is also projected to a pairwise network, in order to evaluate the performances of NS and PD. 
The results are shown in Fig. \ref{fig_sim_matrices_resh_LH10}a-d.
\modified{HPD appears as the most efficient metric for grouping hypergraph instances according to their reshuffling method (Fig. \ref{fig_sim_matrices_resh_LH10}b), whereas 
the separation between instances of the PS and DS methods is visually less clear.}

Table \ref{tab:RI_DI_reshuff_LH10} gives more insights through 
the values of the Rand and Dunn indices obtained
when using each matrix of dissimilarities to cluster the hypergraph instances in three groups,
using the same clustering algorithms as in the previous Section.
\modified{Akin to the case of the clustering of hypergraph models, HNS, HPD, and NS are accurate in finding the ground truth clusters, yielding a Rand Index equal to 1.
On the other hand, PD cannot recover the ground truth groups in case of the ``minimum" algorithm, and provides a good classification only with the ``average" method ($RI=0.99$).
The values of the Dunn Index show that the higher-order metrics produce a larger separation between clusters with respect to NS, in particular HPD, that in this case outperforms the other measures ($DI=2.49$).
Further results showing the superior performances of HNS and HPD with respect to NS and PD are reported in the SI, where we repeat the previous analysis for other data sets.
}

As an additional illustration of the performances of the four metrics, we compute and show in 
Fig. \ref{fig_sim_matrices_resh_LH10}e the average dissimilarity between the original hypergraph
and its randomized versions. Intuitively, the dissimilarity should be smaller for ``stricter"
randomizations, i.e. randomizations that preserve more properties of the original hypergraph.
Figure \ref{fig_sim_matrices_resh_LH10}e
shows that this is indeed the case, as all metrics yield a larger dissimilarity for the RS method, which preserves only the statistics of hyperedge sizes.
\modified{The lowest dissimilarity values are recovered for the DS method for every measure, with a clear distinction between values obtained by the DS and RS methods.}

\begin{table}[t!]
    \begin{ruledtabular}
        \begin{tabular}{ccccc|cccc}
             &\multicolumn{4}{c}{minimum} \vline &\multicolumn{4}{c}{average} \\ \hline
            &HNS &HPD &NS &PD &HNS &HPD &NS &PD\\ \hline
             RI &1.00 &1.00 &1.00 &0.78 &1.00 &1.00 &1.00 & 0.99 \\
            DI &0.96 &2.49 &0.67 &0.52 &0.96 &2.49 &0.67 &0.48 \\
        \end{tabular}
    \end{ruledtabular}
    \caption{Rand Index ($RI$) and Dunn Index ($DI$) for the clustering of randomized hypergraphs (see Fig. \ref{fig_sim_matrices_resh_LH10}). The clusters can depend both on the dissimilarity metric (HNS, HPD, NS, PD) and on the algorithm (``minimum" or ``average") considered.}
    \label{tab:RI_DI_reshuff_LH10}
\end{table}

The results presented above are dependent on the data set from which the reference hypergraph is built and on its structure. For example, 
if in the original system there are no correlations between nodes and specific sizes of interaction, it is likely that the PS and the DS randomization methods will yield hypergraphs that do not differ much from each other. We thus show results with other empirical data sets in the SI, obtaining that HNS and HPD \modified{generally} perform either as well or better than their pairwise counterparts NS and PD.
\modified{Furthermore, in the SI we also analyze the weighted projection of the randomized versions of the LH10 data set. We find results analogous to the ones obtained for the hypergraph models: even though the Rand and Dunn indices of NS and PD slightly increase, the best clustering is still given by the higher-order metrics.
}

\subsection{Clustering of real-world hypergraphs}
We now apply our metrics to several empirical hypergraphs by considering several data sets describing systems belonging to different contexts characterized by higher-order interactions. 
We examine data of face-to-face interactions between individuals, collected by different collaborations and in various environments, such as schools (Utah \cite{toth2015}, Thiers13, LyonSchool \cite{sociopatterns, genois2018}), conferences (SFHH \cite{sociopatterns, genois2018}, ECIR19, ECSS18 \cite{Genois2023}), workplaces (InVS13, InVS15 \cite{sociopatterns, genois2018}), a university (CopNS \cite{Sapiezynski2019}) and a hospital (LH10 \cite{Vanhems2013}). 
We also consider a data set of scientific collaborations (APS \cite{aps_data}) that describes coauthorships in various journals (PRA, PRB, PRC, PRD, PRE, PRL) aggregated on a time window of $5$ years (generally 1992-1996), and data sets of online interactions, such as reviews of products (music-reviews \cite{benson_data, ni2019}) or opinion exchanges in scientific forums (algebra-questions, geometry-questions \cite{benson_data, amburg2020}).
Finally, we consider data describing political interactions in the U.S. Congress (house-committees, senate-committees \cite{benson_data, chodrow2021, stewart2017}). These data sets cover a wide range of system sizes and interaction sizes (see Methods for a detailed description of each data set).

In the previous Sections, we have considered clustering tasks for which a prior knowledge on the right grouping (ground truth) was available: this made it possible to measure the quality of the clustering obtained when using each dissimilarity metric 
-- and hence to assess the significance of the underlying metric itself.
This is not possible when dealing with real-world systems, as generally we have no control on the mechanisms that generate them, and network or hypergraph representations of very different systems
might share non-trivial properties \cite{barabasi2012,newman_book,barrat2008dynamical,bick2023,battiston2020,mancastroppa2023hyper}.
Nevertheless, we can reasonably expect hypergraphs describing analogous systems (or collected with similar techniques) to be more similar to each other than hypergraphs representing systems of different nature.
To understand whether this intuition is confirmed by our metrics, 
we compute the dissimilarity matrices obtained by computing the HNS and HPD 
between all pairs of empirical hypergraphs (or projected \modified{networks} for the NS and PD metrics)
and perform an agglomerative clustering with the \modified{``minimum"} method discussed in the previous Sections (see SI for the results obtained with the \modified{``average"} method). 

\begin{figure*}[hbt!]
    \includegraphics[width=\textwidth]{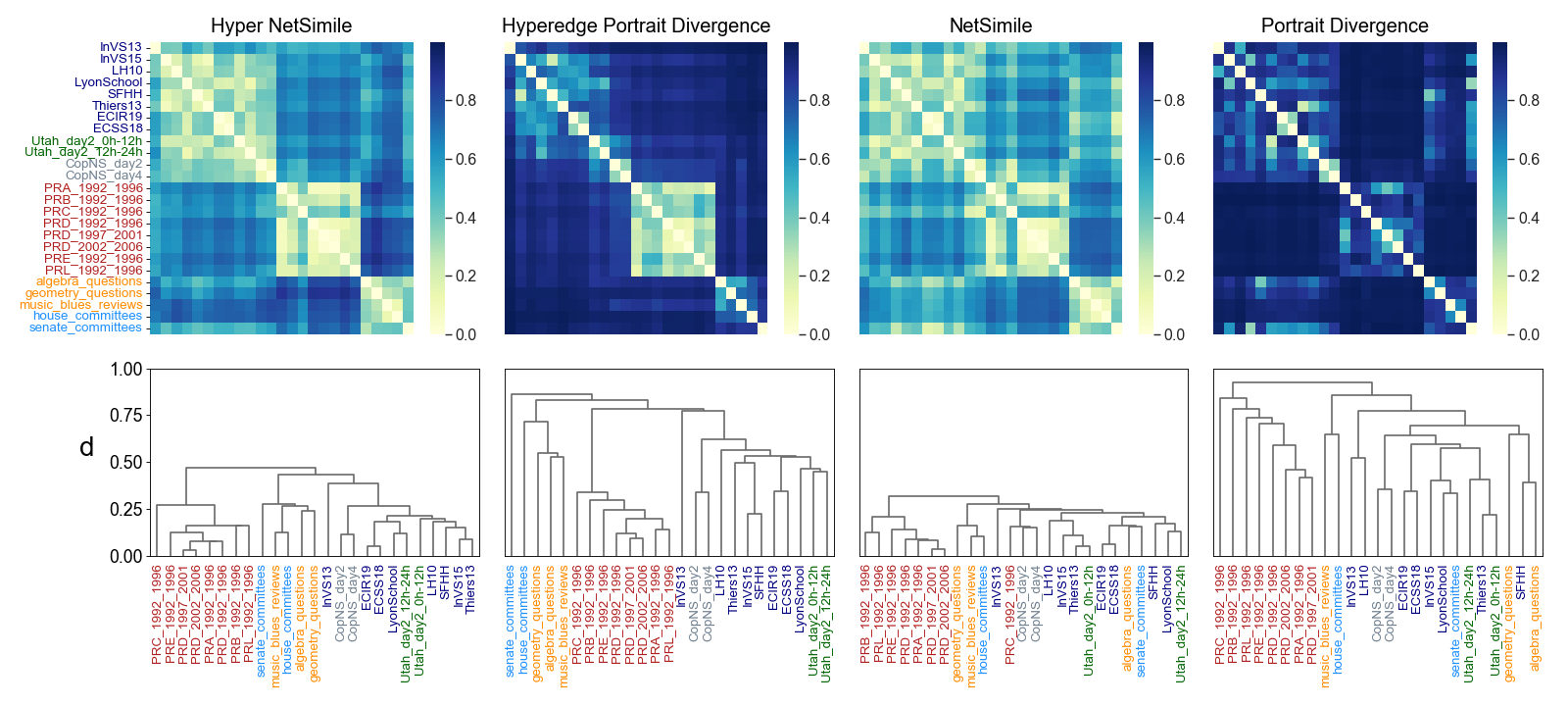}   \caption{\label{fig_sim_matrices_dendro_data} \textbf{Clustering of empirical hypergraphs.} First row: dissimilarity matrices obtained with higher-order (HNS, HPD) and pairwise (NS, PD) metrics. Second row: dendrograms given by the clustering performed with the \modified{``minimum"} agglomerative clustering algorithm.
    The colors of the labels reflect the type of data from which the hypergraphs are built: red for co-authorship, yellow for online interactions, light blue for committees membership; all the remaining labels indicate face-to-face interactions data sets and the colors reflect the different collaborations that collected the data (blue for SocioPatterns \cite{sociopatterns}, gray for the Copenhagen Network Study \cite{Sapiezynski2019}, and green for Utah’s School-age Population project \cite{toth2015}).}
\end{figure*}

Figure \ref{fig_sim_matrices_dendro_data} displays the dissimilarity matrices and the corresponding dendrograms obtained by the clustering algorithm.
Both show that the clusters resulting from higher-order measures are better aligned with the prior knowledge on the data sets than the clusters obtained through pairwise measures.
First, the co-authorship hypergraphs are clearly grouped together by all the metrics, with the exception of NS that misclassifies one of them (PRC\_1992\_1996), placing it closer to some hypergraphs of face-to-face interactions.
Furthermore, according to HNS, HPD, and NS (but not PD), \modified{the two most similar elements among the APS hypergraphs are the ones built from data corresponding to the same journal (PRD) in consecutive periods of time (1997-2001, 2002-2006).}
This is an indicator of the reliability of the metrics, as different fields of research may display a different structure of scientific collaborations \cite{Newman2001,Mancastroppa2024}.
The hypergraphs built from online and political interactions appear to be quite similar to each other according to \modified{HNS, HPD and NS}. Only HPD is able to separate them, 
although not in a clear way. 
\modified{This suggests that these two types of data sets may actually share some structural similarities although they represent different kinds of interactions.}
We also note that NS and PD tend to mix the online and Congress data sets not only with each other, but also with several graphs describing social interactions corresponding to physical proximity.
\modified{On the other hand, HNS and HPD can effectively separate the group of 
face-to-face interaction hypergraphs from the rest
}
Finally, the two hypergraphs built from the Copenhagen Network Study (CopNS) data set are very similar to each other according to all the metrics. 
This fact may reflect the different technique with which the data were collected: CopNS was based on Bluetooth signals of cellphones to detect proximity among individuals (not necessarily corresponding to very close proximity) \cite{Sapiezynski2019}, while the other data sets were collected through RFID wearable proximity sensors able to detect close face-to-face proximity \cite{sociopatterns, genois2018}.

Overall, the higher-order dissimilarity measures provide a clustering corresponding better to the different nature of the empirical data sets than metrics based only on pairwise representations of the data.
Moreover, these findings appear to be robust with respect to the choice of the clustering algorithm (see SI).

\section{Discussion}
We have here introduced two dissimilarity measures for comparing higher-order networks, namely Hyper NetSimile and Hyperedge Portrait Divergence. The former leverages local structural features, with a node-centric point of view, while the latter relies on the statistics of 
paths connecting hyperedges of different sizes, 
and adopts a hyperedge-centric \modified{perspective}.
Both measures are invariant under relabeling of nodes and hyperedges, moreover they do not require any correspondence between the nodes of the systems to be compared, making it possible to consider any arbitrary pair of hypergraphs and in particular hypergraphs of different sizes.
We have shown that their ability to take into account group interactions, going beyond standard network representations, allows for a better distinction and classification of interconnected systems.
This result holds true for both model-generated and real-world hypergraphs, showing a superior performance of the proposed higher-order measures compared to the pairwise ones, which
can only be applied on projected networks in which higher-order features are lost.
Our findings underline the importance of higher-order connections not only as a key aspect for the dynamics
taking place in networked systems \cite{iacopini2019, skardal2020,ferrazdearruda2023},
but also as a relevant structural signature \cite{mancastroppa2023hyper,Mancastroppa2024}, which
can be exploited for identifying and classifying such systems.

Quantitative measures to compare higher-order networks can moreover find applications in several contexts.
Dissimilarity measures can be useful to evaluate the performance of methods to generate synthetic data.
For instance, when generating surrogate data supposed to mimic an empirical data set, the proposed metrics can be used as a model validation tool, to check whether the synthetic objects are sufficiently similar to the empirical ones, thus giving information about the reliability of the generative model.
Analogously, dissimilarity measures can assess the quality of methods for reconstructing hypergraphs from the observation of time series data. In this case, the aim is to infer the underlying set of interactions from the temporal behavior of some observables.
A possible way to check whether the reconstruction method is efficient, is to start with a ground-truth hypergraph, simulate a dynamical process on top of it, and then try to reconstruct the original structure from the observational data.
The quality of the reconstruction can be evaluated by measuring the similarity between the ground-truth hypergraph and the reconstructed one \cite{Young2021, Lizotte2023}.
Finally, a natural application concerns the analysis of time-varying systems. 
In the time-varying graphs framework, tools have been developed to detect the temporal states of a system by comparing snapshots of the temporal network representing it \cite{masuda_holme2019}.
Our metrics makes it possible to extend these approaches to those systems that are better described by temporal hypergraphs, rather than temporal networks \cite{Mancastroppa2024}.

\modified{
The scaling of the time-complexity with the size of the considered hypergraphs is crucial for the actual implementation of the proposed measures.
The computational effort required by HNS is limited, as this metric is based only on local features, such as hyperdegree and hyperedge sizes, that are generally fast to compute.
On the other hand, the computational cost of Hyperedge Portrait Divergence is $\mathcal{O} \big( E + N( \langle k^2\rangle - \langle k\rangle ) \big)$, where $N$ is the number of nodes in the hypergraph, $E$ is the number of hyperedges, $\langle k \rangle$ is the average hyperdegree, and $\langle k^2 \rangle$ is the average squared hyperdegree (see SI for further details).
This may represent a limitation to the application of HPD to large systems, although the results presented here did not require an excessive computational cost and were obtained on a standard laptop computer.
Scaling to very large hypergraphs would probably require some approximations, such as sampling over the set of hyperedges in order to speed up the computation of the hyperedge portrait \cite{madkour2017}.
Assessing the performance of such approximate dissimilarity measure represents an interesting avenue for future work.
}

We finally note that alternative approaches to quantify the similarity between hypergraphs could involve spectral analysis.
In the context of pairwise networks several spectral methods have been proposed, typically based on the eigenvalues of the Laplacian matrix associated with the network
\cite{shimada2016}. However, for higher-order systems there are many possible choices of Laplacian 
operator, corresponding to different notions of diffusion. Defining Laplacian-based similarity measures
between hypergraphs is thus not straightforward \cite{Lucas2020, horak2013, schaub2020, nurisso2025}, even if it represents an interesting research direction.

To conclude, our study provides some practical tools to better characterize and compare systems that are well represented by hypergraphs. Given the increasing relevance of higher-order networks, we expect the proposed methods to be useful in the study of a broad range of complex systems.

\section*{\label{Methods}Methods}

\subsection*{\textit{Data description and preprocessing}}
The data sets we consider are publicly available and describe a wide range of systems represented by hypergraphs.

\paragraph*{Face-to-face interactions.} 
Eight data sets were collected by the Sociopatterns collaboration \cite{sociopatterns, genois2018,Genois2023} and one by the Utah’s School-age Population project \cite{toth2015}.
In both cases the data were obtained through RFID wearable proximity sensors to record face-to-face interactions in several environments: two workplaces (InVS13, InVS15 \cite{genois2015}), three conferences (SFHH, ECIR19, ECSS18 \cite{Genois2023}), a hospital (LH10 \cite{Vanhems2013}), one high school (Thiers13 \cite{Mastrandrea2015}), and two primary school (LyonSchool \cite{Stehle2011}, Utah \cite{toth2015}).
Since these data sets include temporal information on the interactions (i.e., they are represented by temporal networks, with a time resolution of 20 seconds), we preprocess them as follows to obtain static hypergraphs.
First, we aggregate the data over time windows of 15 minutes; then we identify the maximal cliques in each time window, i.e. groups of nodes forming a fully connected cluster, and convert them to hyperedges \cite{mancastroppa2023hyper}.
For all the data sets we consider the full time span of data collection, except for the Utah school, where we consider only the second school day and divide it in two temporal windows of 12 hours each, generating two separate hypergraphs.

We also consider time-resolved data describing physical proximity between students in a university, collected through the Bluetooth signal of cellphones within the Copenhagen Network Study \cite{Sapiezynski2019}. We restricted our analysis to two days (day2 and day4, i.e., the Monday and Wednesday of the first week of data collection) and preprocessed the data aggregating over time windows of 5 minutes, as described in \cite{Iacopini2024,Mancastroppa2024}.

\paragraph*{Co-authorship.} 
The American Physical Society (APS) scientific collaborations data set \cite{aps_data} includes the APS publications from 1893 to 2021. For each paper the date of publication, the journal and the list of authors are indicated.
We consider six journals (PRA, PRB, PRC, PRD, PRE, PRL) and we preprocess the data as described in \cite{Mancastroppa2024}.
We then build the hypergraphs in which each node is an author and a hyperedge represents a paper connecting the co-authors, published in the corresponding journal. We focus on temporal windows of 5 years (1992-1996) for all the journals and we also consider the periods 1997-2001 and 2002-2006 for PRD, in order to check whether the similarity among these co-authorship hypergraphs is affected by the choice of the temporal period.

\paragraph*{Online interactions.} 
We consider two data sets describing exchanges between users of MathOverflow on algebra topics (algebra-questions) or on geometry topics (geometry-questions). Each node corresponds to a user of MathOverflow and each hyperedge involves those users who have answered a specific question belonging to the topic of algebra or geometry \cite{benson_data, amburg2020}. The third data set represents the interactions between Amazon users on music (music-review \cite{benson_data,ni2019}), in which each node corresponds to an Amazon user and each hyperedge involves users who have reviewed a specific product belonging to the category of blues music.

\paragraph*{Political interactions.} 
We consider data describing the interactions in committees in the U.S. House of Representatives (house-committees) and in the U.S. Senate (senate-committees) \cite{benson_data, chodrow2021, stewart2017}. Each node corresponds to a member of the U.S. House of Representatives or Senate and each hyperedge involves nodes that share membership in a committee.

\subsection*{\textit{Generative models of hypergraphs}}
In the following we specify the details about the generative models that we introduced in the main text.

\paragraph*{Erd\H{o}s-Rényi (ER).}
This model is the higher-order generalization of the Erd\H{o}s-Rényi graph \cite{Bollobas2001, newman_book}; it has generally as many parameters as the number of orders of interaction. We can equivalently specify either the probability $p_s$ of creating an hyperedge involving $s$ nodes chosen at random, or the expected $s$-degree of nodes $\langle k_s\rangle$. 
Here, we consider the average degree to be the same for all sizes of interactions, i.e. $\langle k_s\rangle = \bar{k}\ \ \forall s $. 
\modified{Then, a value of $\bar{k}$ is drawn from a uniform distribution in the interval $[1.5,2]$ for each realization of the model. 
}
\modified{The hypergraph is built layer by layer, i.e. size by size, up to a hyperedge size of $7$},
considering the orders of interaction independently from each other and selecting the nodes in each hyperedge uniformly at random.

\paragraph*{Configuration Model (CM).}
This model is the higher-order generalization of the configuration model of graphs \cite{Bollobas2001, newman_book}; it generates a hypergraph starting from a given sequence of $s$-hyperdegrees, replicating it. 
This means that the user can specify the degree $k_s(i)$ of each node $i$ at each order of interaction $s$. 
This might lead to configurations that are not realizable, unless self-loops or multiple edges are allowed. To solve these cases, the algorithm we used slightly increases the degree of random nodes until a realizable sequence is obtained. The shape of the degree distributions is however not affected by these small possible variations. 
In order to simplify the procedure, for each $i$ we extract 
a random value $k(i)$ from a fixed degree distribution and use
the same degree value for all sizes $s$:
$k_s(i)=k(i)\ \ \forall s$. 
This lowers the number of parameters of the model to $N$, that is the number of nodes in the hypergraph.
To build the sequence of degrees, we draw $N$ samples from a power-law distribution $P(x) = (\gamma-1) x^{-\gamma}$ defined on the interval $x\in [1,+\infty]$ and take the degree of node $i$ as $k(i) = \min (x_i,30)$.
For each instance of the model, the exponent $\gamma$ is randomly chosen in the interval \modified{$[2,2.05]$}. As in the Erd\H{o}s-Rényi model, we build the hypergraph in a stratified way by considering one size of hyperedges at time, \modified{up to size $3$}.

\paragraph*{Watts-Strogatz (WS).}
Here we propose a higher-order generalization of the Watts-Strogatz model \cite{WS}, once again built in a stratified manner \modified{up to size $5$}. 
The easiest way to think of it is in terms of the combinatorial sequence of the hyperedges. 
Let us consider the order of interaction $s$ and assume that the nodes are labeled from 1 to $N$. We create an hyperedge among the first $s$ nodes, then shift by one the labels of the nodes and create another hyperedge including the nodes $2,...,s+1$. We iterate the process until we connect the last nodes with the first ones. 
For example, for $s=2$ we have the edges $(1,2),(2,3),(3,4),...,(N,1)$, while for $s=3$ the sequence of hyperedges is $(1,2,3),(2,3,4),(3,4,5),...,(N,1,2)$.
Once we have put together the sets of interactions of different sizes, every hyperedge is rewired with probability $p_{rew}$. In this case, a randomly chosen node remains in the selected hyperedge, while the other $s-1$ nodes are replaced by elements selected at random among the other $N-s$ possible nodes.
Therefore, $p_{rew}$ is the only free parameter that has to be set. For each realization of the model we draw $p_{rew}$ from a uniform distribution in the interval \modified{$[0.15,0.2]$}.

\subsection*{\textit{Hypergraph randomization methods}}
In the main text we employ the following randomization methods to reshuffle the hyperedges within a hypergraph.
\paragraph*{Random Shuffling (RS).}
This routine preserves only the number and size of the hyperedges, while it destroys any other property of the system.
All the nodes in every $s$-hyperedge of the original hypergraph are replaced by other $s$ nodes randomly chosen within the $N$ possible.
Most of the structural properties are lost in this procedure, such as correlations between nodes and sizes of interactions, meso-scale structure, and hyperdegree distribution and correlations.

\paragraph*{Proportional Shuffling (PS).}
Akin to the former method, the proportional shuffling reassigns $s$ nodes to every $s$-hyperedge of the original hypergraph, but the nodes are chosen with probability proportional to their hyperdegree.
This means that a node taking part in many hyperedges in the original hypergraph is likely to do so also in the randomized system.
However, this method is not sensitive to the specific sizes of interaction in which a node is involved, as the hyperdegree accounts just for the total number of hyperedges a node is part of. 
Thus, although we expect the total hyperdegree distribution to remain approximately unchanged, if we look order by order (i.e. the $s$-degree distributions) some differences may arise between the original and the randomized hypergraph.
Furthermore, this method still destroys meso- and large-scale structures and correlations between nodes and specific sizes of interaction.

\paragraph*{Degree-preserving Shuffling (DS).}
With this method we aim to preserve exactly the $s$-degree of every node and for each value of $s$.
We do so through a double-edge swap. First, two hyperedges $e_1$ and $e_2$ with same size $s$ are randomly selected in the original hypergraph. Then, we choose at random a node $n_1$ within $e_1$ and a node $n_2$ within $e_2$. Finally, we swap the membership of $n_1$ to $e_1$ with the one of $n_2$ to $e_2$, obtaining $e_1^{new} = (e_1 \smallsetminus n_1 ) \cup n_2$ and $e_2^{new} = (e_2 \smallsetminus n_2 ) \cup n_1$. The other properties of $n_1$ and $n_2$ remain unchanged.
By iterating this process and applying it to every order of interaction we erase once again the possible community structure of the hypergraph, as well as the correlations between different orders of interaction. However, the $s$-degree distributions and the correlations between nodes and sizes of interaction are preserved through this reshuffling method \cite{mancastroppa2023hyper}.

\subsection*{\textit{Rand and Dunn indices}}
The Rand Index ($RI$) and the Dunn Index ($DI$) quantify the quality of a clustering method.
The $RI$ can be applied to compare the output of a clustering algorithm with a grouping that is known to be the correct one.
It measures the correspondence between the ground truth labels of the elements and the predicted ones, assigned by the clustering algorithm. Formally, it reads:
\begin{equation}
    RI = \frac{a+b}{\binom{\mathcal{N}}{2}},
    \label{RI}
\end{equation}
where $a$ is the number of pairs of elements that belong to the same class and are assigned to the same cluster by the algorithm, $b$ is the number of pairs of elements that belong to different classes and are assigned to different clusters by the algorithm, and $\mathcal{N}$ is the total number of elements.
The $RI$ takes values between 0 and 1, and the maximum is reached only if all the elements are divided in the correct classes.

The Dunn Index ($DI$), on the other hand, does not rely on a ground truth grouping. 
Given a clustering algorithm, it considers the groups in which the elements are divided and measures the degree of separation between clusters. It is defined as:
\begin{equation}
    DI = \frac{\min_{c,c' \in C,\ c\neq c'}\ \min_{i\in c,\ j\in c'} \ d(i,j)}{\max_{c\in C} \ \max_{i,j \in c}\ d(i,j)}\ ,
    \label{DI}
\end{equation}
where $C$ is the set of clusters given by the algorithm. 
In other words, the right-hand side of Eq. (\ref{DI}) is given by the minimum inter-clusters distance divided by the maximum intra-cluster distance. 

\section*{Data availability}
The data that support the findings of this study are publicly available. The APS data set can be requested at \url{https://journals.aps.org/datasets}; the SocioPatterns data sets are available at \url{http://www.sociopatterns.org/}, and
at \url{https://search.gesis.org/research_data/SDN-10.7802-2351?doi=10.7802/2351} for the conference
data sets; 
the online and political interactions data sets at \url{https://www.cs.cornell.edu/~arb/data/}; 
the Contacts among Utah’s School-age Population data set at \url{https://royalsocietypublishing.org/doi/suppl/10.1098/rsif.2015.0279}; 
the Copenhagen Network Study data set at \url{https://doi.org/10.6084/m9.figshare.7267433}.

\section*{Code availability}
The code to reproduce the results of this work is available at \url{https://github.com/cosimoagostinelli/Hor_dissimilarity_measures}.
The code is based on the Python package XGI \cite{XGI}, an open-source library that provides tools to manage the structure and dynamics of higher-order networks.

\section*{Acknowledgments}
M.M. and A.B. acknowledge support from the Agence Nationale de la Recherche (ANR) project DATAREDUX (ANR-19-CE46-0008).
C.A. and A.B. acknowledge support by the “BeyondTheEdge: Higher-Order Networks and Dynamics” project (European Union, REA Grant Agreement No. 101120085).

\section*{Authors' contributions}
A.B. conceptualized the study. 
C.A. developed the algorithms and performed the numerical experiments. 
A.B., C.A. and M.M. analyzed the results and contributed to writing the manuscript.

\section*{Competing interests}
The authors declare no competing interests.

\bibliographystyle{naturemag}
\bibliography{references_noURL}

\end{document}


\title{Supplementary Information for ``Higher-order dissimilarity measures for hypergraph comparison"
}

\author{Cosimo Agostinelli}
\affiliation{Aix-Marseille Univ, Université de Toulon, CNRS, CPT,
Turing Center for Living Systems, 13009 Marseille, France}
\author{Marco Mancastroppa}
\affiliation{Aix-Marseille Univ, Université de Toulon, CNRS, CPT,
Turing Center for Living Systems, 13009 Marseille, France}
\author{Alain Barrat}
\affiliation{Aix-Marseille Univ, Université de Toulon, CNRS, CPT,
Turing Center for Living Systems, 13009 Marseille, France}

\renewcommand*{\citenumfont}[1]{S#1}
\renewcommand*{\bibnumfmt}[1]{[S#1]}

\setcounter{figure}{0}
\setcounter{table}{0}
\setcounter{equation}{0}
\setcounter{page}{1}
\setcounter{section}{0}

\makeatletter
\renewcommand{\thepage}{\roman{page}}
\renewcommand{\thefigure}{S\arabic{figure}}
\renewcommand{\theequation}{S\arabic{equation}}
\renewcommand{\thetable}{S\arabic{table}}

\setcounter{secnumdepth}{2}

\maketitle

\onecolumngrid

In this Supplementary Information we present some additional results: 
we present the results for the clustering of hypergraph models when skewness and kurtosis are taken into account in the computation of HNS and NS and also when some features related to the clustering coefficient are added to HNS;
we investigate the ability of the different proposed metrics to distinguish hypergraph instances of the ER model generated with different parameters; 
we consider the performance of NS and PD metrics on weighted graphs obtained by projecting hypergraphs by retaining more information from their higher-order properties;
we show the efficiency of the four metrics in distinguishing different randomization methods applied to several data sets representative of various contexts, not presented in the main text;
we present the dendrograms associated with the clustering of empirical hypergraphs performed via the ``minimum" clustering algorithm;
finally, we provide an argument to estimate the computational complexity of HPD.

\section*{Alternative formulation of Hyper NetSimile and NetSimile}

\begin{figure*}[t!]
    \includegraphics[width=\textwidth]{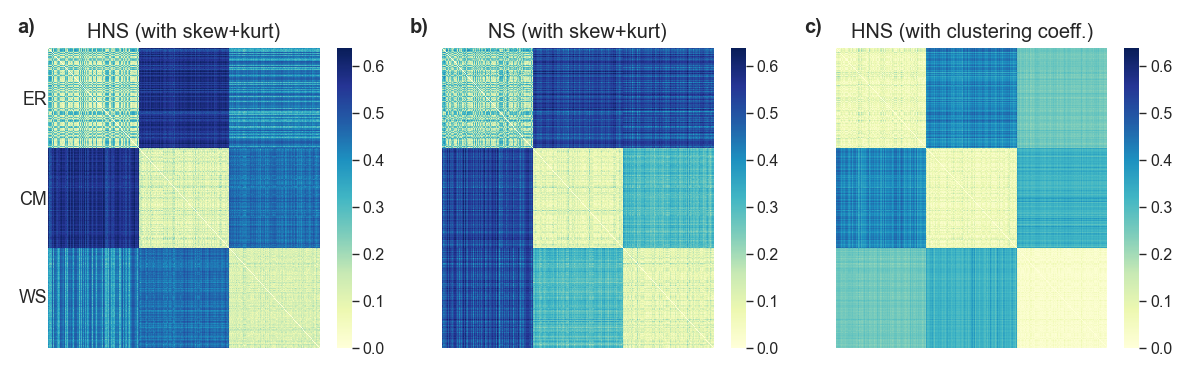}
    \caption{\label{SI_fig_skew_kurt_clst} \modified{\textbf{Dissimilarity matrices of hypergraph models, computed with alternative definitions of HNS and NS.} \textbf{(a-b)} Dissimilarity matrices for HNS and NS when skewness and kurtosis are included among the statistical indicators for building the feature vector.
    \textbf{(c)} Dissimilarity matrix for HNS when two more features related to the clustering coefficient are added to the summary vector representing the hypergraph.
    }}
\end{figure*}

\modified{We consider the original formulation of NetSimile \cite{Berlingerio2012} that includes also skewness and kurtosis among the statistical indicators used to build the feature vector of a network.
Similarly, we include these two additional indicators also in the formulation of Hyper NetSimile.
We test this version of NS and HNS on the same clustering task of hypergraph models presented in the main text (see Section ``Clustering of hypergraph model instances"), obtaining a Rand Index equal to 1 for both measures, as in the case presented in the main text.
However, the Dunn Index of HNS drops from $DI=1.48$ to $DI=0.62$ when including skewness and kurtosis, and the one of NS decreases form $DI=0.78$ to $DI=0.40$, thus indicating a worse separation between clusters.
These values are obtained with both the ``minimum" and the ``average" clustering algorithm, and motivate the choice of omitting these two statistical indicator in the definition of HNS.
The corresponding dissimilarity matrices are shown in Fig. \ref{SI_fig_skew_kurt_clst}a-b.
}

\modified{We also investigate the impact of the clustering coefficient on the performance of HNS, when it is included in the list of features extracted from the hypergraph.
We consider the definition of hyper clustering coefficient presented in \cite{Zhou2011}, that generalizes the standard notion of clustering to the case of hypergraphs.
We add two more distributions to the list of seven features presented in the main text, namely the hyper clustering coefficient of the nodes and the average hyper clustering coefficient of each node's neighbors.
Again, we test this extended version of HNS on the clustering task of hypergraph models, as presented in the main text, and show the dissimilarity matrix in Fig. \ref{SI_fig_skew_kurt_clst}c.
We find that the Rand Index remains equal to 1, while the Dunn Index drops from $DI=1.48$ to $DI=0.99$.
Thus, we conclude that it is possible to avoid using the hyper clustering coefficient when building the summary vector of a hypergraph, without losing the performance of HNS.
This finding turns out in fact to be very useful, as the computational cost of the hyper clustering coefficient is very high, whereas none of 
the other features present such a computational cost issue.
This makes HNS a suitable dissimilarity measure for scaling to large hypergraphs.
}

\section*{\label{SI_ER_models}Clustering of ER models with different maximum hyperedge size}

Let us consider the higher-order Erd\H{o}s-Rényi model presented in the main text.
We want to check whether the four metrics considered (two higher-order and two lower-order) can distinguish among different realizations of the model when varying the parameters.
In particular, here we focus on the maximum hyperedge size $M$, as it is a relevant property for characterizing many real-world hypergraphs \cite{iacopini2024_SI}.
We thus consider the ER model and for each realization we choose the number of nodes uniformly at random in the interval $[200, 300]$. We consider three cases, namely $M=3,\ M=4,\ M=5$, and set the probabilities $p_s$ of creating an $s$-hyperedge in such way as to keep the average projected degree close to \modified{$\langle k_{prj} \rangle = 20$}, for any value of $M$ (see the paragraph \textit{Implementation of the ER models} below). This means that if we consider the pairwise projection $\mathcal{G(H)}$ of any of the instances $\mathcal{H}$ of the model, it will have an average degree close to 20, regardless of the value of $M$.
The rationale behind this constraint is that we want to explore scenarios where the differences between networked systems are mostly higher-order.

\begin{figure*}[t!]
    \includegraphics[width=\textwidth]{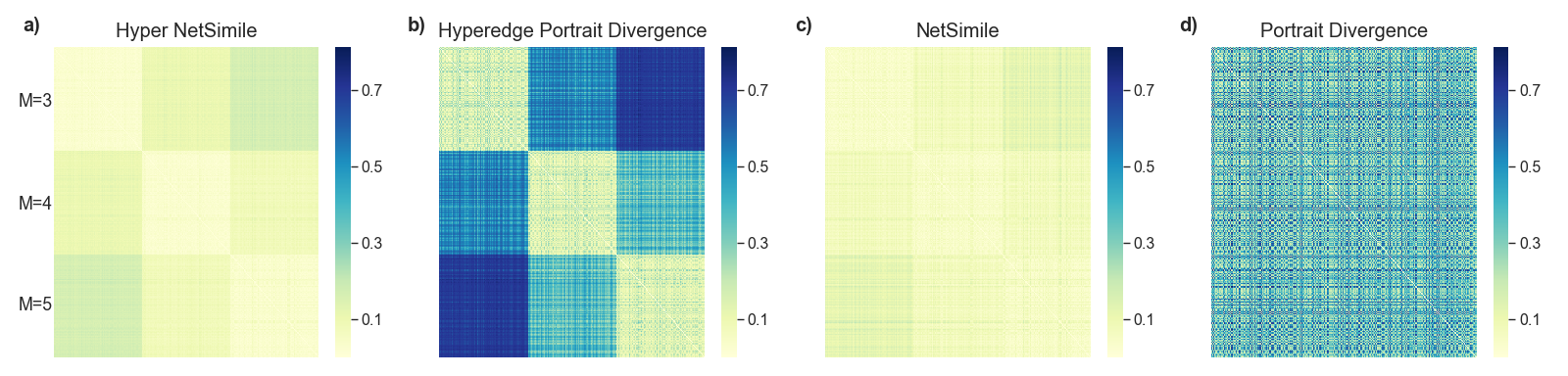}
    \caption{\label{fig_sim_matrices_ERs_3_4_5} \textbf{Dissimilarity matrices of ER models with different maximum size of hyperedges.} The size is indicated with $M=3,\ M=4,\ M=5$ and the average projected degree is fixed to $\langle k_{prj} \rangle = 20$. The dissimilarity matrices are computed with higher-order \textbf{(a-b)} and pairwise metrics \textbf{(c-d)}. For every value of $M$ we consider $100$ instances of the model, with a number of nodes chosen uniformly at random in the interval $[200,300]$.}
\end{figure*}

For our analysis we sample 100 realizations of the ER model for each value of $M$. The $300\times 300$ dissimilarity matrices computed with HNS, HPD, NS, and PD are reported in Fig. \ref{fig_sim_matrices_ERs_3_4_5}, while the values of the Rand Index are shown in Table \ref{tab:RI_models_ER_3_4_5}.
\modified{As expected, in this case NS and PD are not able to retrieve the correct groups, as they feature a low Rand Index. Indeed, at the pairwise level, the hypergraphs considered do not differ much from each other, as they all feature the same random structure.
On the other hand, HNS and HPD provide instead the correct clustering of the hypergraphs even though they are generated by the same model (ER).
This suggests that the higher-order metrics are sensitive not only to the mechanism generating the connections among nodes but also to the diversity of sizes in which the nodes are involved.
}

\paragraph*{Implementation of the ER models.}
To build a set of ER models with variable maximum hyperedge size $M$ and fixed average projected degree $\langle k_{prj} \rangle$, we need to express $p_s$ as a function of $\langle k_{prj} \rangle,\ N,\ M$.
We start noticing that every hyperedge of size $s$ in a hypergraph contributes with $\binom{s}{2}$ edges in the projected network $\mathcal{G(H)}$: this holds if we assume to deal with sparse random hypergraphs (i.e. with $p_s$ small $\forall s$), where we can neglect the effect of the overlap between hyperedges, that would otherwise lower the number of edges in $\mathcal{G(H)}$.
In this case, the expected number of edges in the projected networks $\mathbb{E}[E_{prj}]$ can be approximately written as a sum over $s$ of the contribution given by the different orders of interaction in $\mathcal{H}$:
\begin{equation}
    \mathbb{E}[E_{prj}] \simeq \sum_{s=2}^M  p_s \binom{N}{s} \binom{s}{2} = \binom{N}{2} \sum_{s=2}^M  p_s \binom{N-2}{s-2},
    \label{SI_eq:E_avg}
\end{equation}
where the term $p_s \binom{N}{s}$ is the expected number of hyperedges of size $s$ in $\mathcal{H}$.
To compute the average degree of the projected network, we need to divide the expected number of edges by $N$ and multiply it by 2 so to double-count every edge:
\begin{equation}
    \langle k_{prj} \rangle \simeq (N-1)\sum_{s=2}^M p_s \binom{N-2}{s-2} \ .
    \label{SI_eq:k_prj}
\end{equation}
Now, we can set the probabilities to $p_s = p_2 /\binom{N-2}{s-2}$ in order to have a reasonable number of $s$-hyperedges for every value of $s$. Indeed, this condition implies that the projected degree of a node is composed by equal contributions coming from the projection of the different orders of interaction.
This fact becomes clear when inserting the condition on the $p_s$ in Eq. (\ref{SI_eq:k_prj}). This leads to the expression of $p_2$ as a function of $\langle k_{prj} \rangle,\ N$, and $M$:
\begin{equation}
    p_2 = \frac{\langle k_{prj} \rangle}{(N-1)(M-1)} \ .
    \label{SI_eq:k_prj2}
\end{equation}
Finally we can use Eq. (\ref{SI_eq:k_prj2}) and the condition $p_s = p_2 /\binom{N-2}{s-2}$ to generate ER hypergraphs with different $N$ and $M$ and fixed average projected degree $\langle k_{prj} \rangle$.
We verify \textit{a posteriori} that this procedure gives the desired result, justifying the assumption of non-overlapping hyperedges for the chosen values of the parameters.

\begin{table}[t!]
    {\begin{tabular*}{0.95\textwidth}
    {@{\extracolsep{\fill}}lllll|llll@{}}
    &\multicolumn{4}{c}{minimum} \vline &\multicolumn{4}{c}{average} \\ \hline
    & HNS & HPD & NS & PD & HNS & HPD & NS & PD \\
    \hline
    RI &1.00 &1.00 &0.34 &0.34 &1.00 &1.00 &0.77 & 0.54 \\
    DI &0.42 &0.60 &0.13 &0.26 &0.42 &0.60 &0.15 & 0.22 \\
    \end{tabular*}}{}
    \caption{Rand Index ($RI$) for the clustering of ER hypergraphs with different maximum hyperedge size $M$ (see Fig. \ref{fig_sim_matrices_ERs_3_4_5}). The clusters are determined both by the algorithm (``minimum" or ``average") and the dissimilarity metric (HNS, HPD, NS, PD) considered.
    \label{tab:RI_models_ER_3_4_5}}
\end{table}

\section*{Weighted projection of hypergraphs}

\modified{In this section we consider the so-called \textit{weighted projection} $\mathcal{W}$ of a hypergraph $\mathcal{H}$, which is a more precise way of reducing a hypergraph to a pairwise network.
The idea here is to convert $\mathcal{H}$ to a graph, where each edge $(i,j)$ is associated with a weight given by the number of hyperedges in which both $i$ and $j$ are interacting together in $\mathcal{H}$.
The goal of this type of projection is to represent the system as a pairwise network without losing too much information about the interactions among the nodes.
In this sense, it is a more refined version of the pairwise projection, which instead is generally a raw approximation of higher-order networks, as shown by the results presented in the main text.
}

\modified{Since we now have to compare weighted networks, some choices regarding the dissimilarity measures have to be made.
Both NS and PD can be extended to weighted graphs.
For NetSimile we have to chose the features of interest for building the summary vector representing $\mathcal{W}$.
A reasonable set of features of each node $i$ is the following:
\begin{enumerate}
    \item number of $i$'s neighbors;
    \item $i$'s strength, i.e., the sum of the weights of the edges involving $i$;  
    \item $i$'s weighted clustering coefficient, as defined in \cite{Barrat2004};
    \item standard deviation of the weights of $i$'s edges;
    \item average number of neighbors of $i$'s neighbors;
    \item average strength of $i$'s neighbors;
    \item number of neighbors of $i$'s ego-net, i.e., number of nodes that are two steps away from $i$.
\end{enumerate}
Note that features 1, 5, and 7 are exactly the same of HNS, while in features 2, 4, 6 we have basically replaced the hyperdegree of the node in $\mathcal{H}$ with its strength in $\mathcal{W}$.
The only exception is feature number 3. In HNS we could choose the average size of $i$'s hyperedges because generally it is not redundant with the features 1 and 2 of HNS, that are the number of neighbors and the hyperdegree of $i$, respectively. 
Indeed, the discrepancy between feature 3 and the ratio between features 2 and 1 in HNS accounts for the overlap between hyperedges involving $i$.
This is not true in the case of a pairwise network, and the average strength of a node is just the ratio between its total strength and its number of neighbors.
To replace this redundant feature, we select the weighted clustering coefficient because it is notoriously an important quantity to consider when studying weighted networks \cite{Barrat2004}.
Furthermore, this feature is also included in the original formulation of NS.
This fact places the weighted formulation of NetSimile, as proposed above, halfway between the unweighted NS and HNS.
}

\modified{Regarding PD, the authors of \cite{Bagrow2019} also proposed a method to compute the portrait of a weighted network. However, this comes with some limitations. 
First, the computational complexity increases when considering weighted networks, due to the fact that the shortest paths have to be computed through the Dijkstra algorithm instead of a breadth-first search \cite{cormen2009}.
Secondly, the shortest paths become real-valued, so that a binning strategy is needed in order to build the portraits of two networks and to compare them coherently. 
We refer the reader to the Supplemental Material of \cite{Bagrow2019} for a discussion of these points.
For our purposes, we adopt the following binning method.
Let $\mathcal{W}_1$,  $\mathcal{W}_2$ be the two weighted networks to compare. We consider the interval ranging from the smallest shortest path to the largest one, among all the shortest paths in $\mathcal{W}_1$ and $\mathcal{W}_2$. 
We then divide this interval into 100 percentiles that will act as bins in the computation of the network portraits of $\mathcal{W}_1$ and $\mathcal{W}_2$.
This means that the index $l$ in $B_{l,k}$ will indicate the percentile in which the shortest paths fall, instead of their precise values.
}

\begin{figure*}[t!]
    \includegraphics[width=\textwidth]{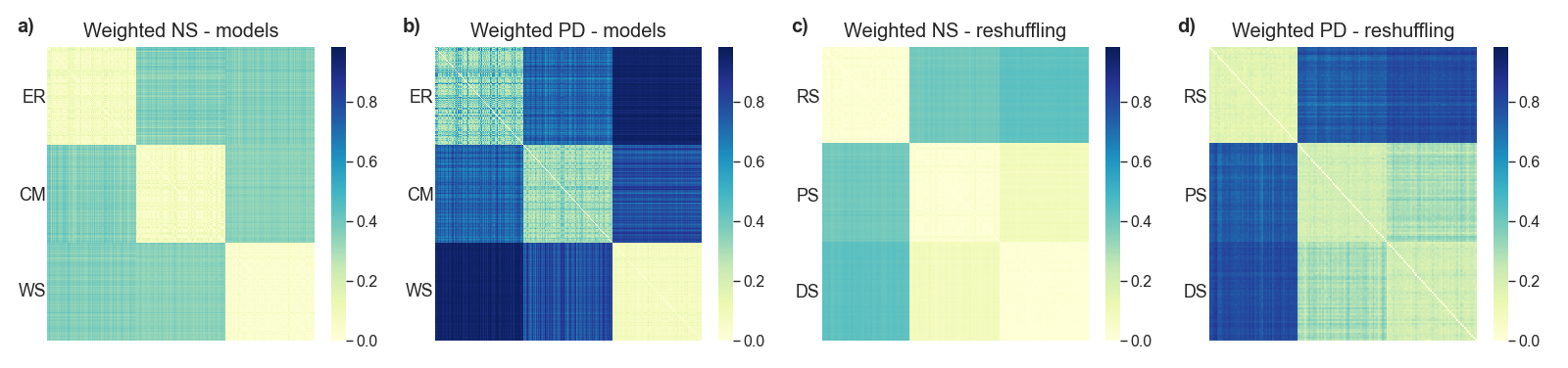}
    \caption{\label{SI_fig_weighted_proj} \modified{\textbf{Dissimilarity matrices of weighted projection of hypergraphs, computed with NS and PD.} \textbf{(a-b)} Dissimilarity matrices of instances of (projected) hypergraph models.
    \textbf{(c-d)} Dissimilarity matrices of (projected) reshuffled hypergraphs, the reference hypergraph being the one built from the SocioPatterns LH10 data set. 
    }}
\end{figure*}

\modified{We test the weighted formulations of NS and PD in two scenarios proposed in the main text.
The first one is the clustering of hypergraphs generated by different models (ER, CM, and WS).
We consider the same realizations of the models analyzed in the main text, and present the dissimilarity matrices in Fig. \ref{SI_fig_weighted_proj}a-b.
The three diagonal blocks are clearly visible for both metrics, and indeed the Rand Index of both NS and PD becomes $RI=1$. This denotes an improvement of PD, which provided only $RI=0.78$ in case of a pairwise projection.
The Dunn Index also increases, with NS and PD yielding respectively
$DI=1.11$ and $DI=0.69$.
These values are found both with the ``minimum" and the ``average" clustering algorithm.
As expected, a weighted projection of the hypergraphs results in a more accurate classification, as a larger amount information about the original structure is retained.
Nonetheless, the higher-order metrics still provide better results than their pairwise (weighted) counterparts, as the Dunn Index for HNS and HPD is $DI=1.48$ and $DI=1.09$ respectively (see main text).
}

\modified{The second case study we consider is the clustering of instances of reshuffling methods.
As in the main text, the starting hypergraph is built from a face-to-face interactions data set, collected in a hospital by the SocioPatterns collaboration \cite{genois2018, sociopatterns, Vanhems2013}.
We repeat the same analysis presented in the main text and show the dissimilarity matrices in Fig. \ref{SI_fig_weighted_proj}c-d. For NS the Dunn Index slightly increases from $DI=0.67$ to $DI=0.75$, the $RI$ remaining equal to 1.
PD maintains instead the same values of $RI$ and $DI$ when the ``minimum" clustering algorithm is applied ($RI=0.78$, $DI=0.52$), while it improves in the case of the ``average" algorithm: $RI=1$, $DI=0.69$, instead of $RI=0.99$, $DI=0.48$ that we had in case of the pairwise projection (see main text).
Yet, also in this case HNS and HPD are still the best metrics to classify the randomized versions of the original hypergraph, providing $RI=1$, $DI=0.96$ for HNS and $RI=1$, $DI=2.49$ for HPD.
}

\modified{Overall, the previous analysis demonstrate that even if a weighted projection retains more information than a simple pairwise approximation, still this is not sufficient to capture all the nuances encoded in higher-order interactions.
Thus, higher-order metrics are generally more suitable to compare systems that are naturally represented by hypergraphs.
}

\section*{\label{SI_reshuffling_methods}Randomization methods}

\begin{figure*}[ht!]
    \includegraphics[width=\textwidth]{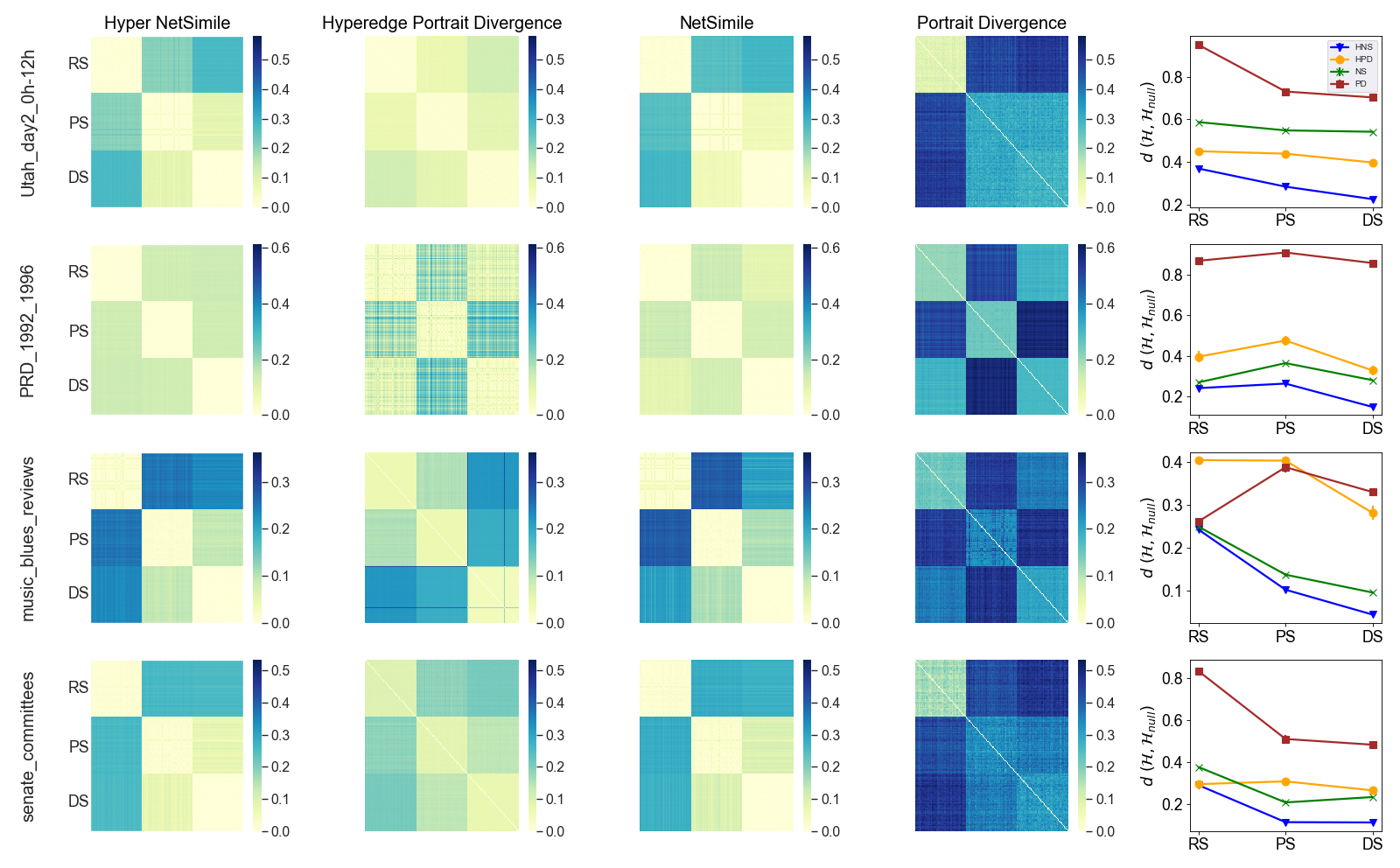}
    \caption{\label{SI_fig_reshuffling_all_data} \textbf{Dissimilarity matrices of randomization methods for several empirical hypergraphs.} We sample 50 realizations of each randomization method (RS, PS, DS) and compute the dissimilarity with higher-order (first two columns) and pairwise metrics (third and fourth column). The last column shows the average dissimilarity between the original hypergraph ($\mathcal{H}$) and the realizations of the three randomization methods ($\mathcal{H}_{null}$), computed with the various measures. The error bars represent the standard deviation.}
\end{figure*}

We repeat the analysis of the randomization methods presented in the main text, applying it to other data sets.
We consider four hypergraphs representing different types of systems: face-to-face interactions (Utah\_day2\_0h-12h \cite{toth2015}), scientific collaborations (PRD\_1992\_1996 \cite{aps_data}), online interactions (music\_blues\_reviews \cite{benson_data, ni2019}), and political interactions (senate\_committees \cite{benson_data, chodrow2021, stewart2017}).
We refer to the Methods in the main text for further details about the data sets.
For each hypergraph we sample 50 realizations of the three randomization methods (RS, PS, DS) and build the dissimilarity matrix for the four metrics (HNS, HPD, NS, PD) 
(Fig. \ref{SI_fig_reshuffling_all_data}, column 1-4). 
We also compute the distance between the original hypergraph and the reshuffled ones (Fig. \ref{SI_fig_reshuffling_all_data}, column 5).
Finally, Table \ref{SI_tab:RI_DI_reshuff_datasets} displays the values of the Rand and Dunn indices corresponding to each dissimilarity matrix for two clustering algorithms (``minimum" and ``average").

\modified{Let us start from the human proximity data set (Utah\_day2\_0h-12h). 
The higher-order metrics retrieve the correct clusters ($RI=1$), while NS and PD cannot distinguish between the PS and the DS reshuffling, as can be seen from the dissimilarity matrices of Fig. \ref{SI_fig_reshuffling_all_data}.
This results in a $RI=0.78$ for NS and PD, with both clustering algorithms.
Notably, HPD features a very high value of the Dunn Index ($DI=4.11$), that indicates a very neat separation between the groups of reshuffled hypergraphs.
Regarding the co-authorship (PRD\_1992\_1996) and the online interactions (music reviews) data sets, only HNS and NS reach a $RI=1$, whereas HPD and PD have lower values of the Rand Index.
This result might indicate that for these particular data sets dissimilarity measures based on paths are not a suitable choice.
In both cases the metric that produces the largest Dunn Index is HNS ($DI=5.65$ for the co-authorship data and $DI=2.04$ for the online interactions data).
Finally, for the committees membership data set, we have that all the metrics give a $RI=1$ except for PD when the clustering is performed via the ``minimum" algorithm.
Here the largest Dunn Index is instead provided by HPD ($DI=1.26$).
Overall, higher-order measures appear as a better tool to differentiate randomization methods, as for every data set the metric with the largest Rand and Dunn indices is always either HNS or HPD, independently on the choice of the clustering algorithm.
}

\begin{table}[b!]
    {\begin{tabular*}{0.95\textwidth}
    {@{\extracolsep{\fill}}l|lllll|llll@{}}
    &\multicolumn{5}{c}{minimum} \vline &\multicolumn{4}{c}{average} \\ \hline
            & &HNS &HPD &NS &PD &HNS &HPD &NS &PD\\ \hline
            \multirow{2}{*}{Utah 0h-12h} &\modified{RI} &1.00 &1.00 &0.78 &0.78 &1.00 &1.00 &0.78 & 0.78 \\
                                         &\modified{DI} &1.21 &4.11 &0.75 &0.77 &1.21 &4.11 &0.75 &0.77 \\ \hline
            \multirow{2}{*}{PRD 1992-1996} &\modified{RI} &1.00 &0.77 &1.00 &0.78 &1.00 &0.87 &1.00 & 0.97 \\
                                           &\modified{DI} &5.65 &0.34 &2.79 &0.88 &5.65 &0.26 &2.79 &0.89 \\ \hline
            \multirow{2}{*}{music reviews} &\modified{RI} &1.00 &0.77 &1.00 &0.77 &1.00 &0.77 &1.00 &0.76 \\
                                           &\modified{DI} &2.04 &1.06 &1.83 &0.68 &2.04 &1.06 &1.83 &0.67 \\ \hline
            \multirow{2}{*}{Senate committees} &\modified{RI} &1.00 &1.00 &1.00 &0.78 &1.00 &1.00 &1.00 & 1.00 \\
                                               &\modified{DI} &0.72 &1.26 &0.97 &0.75 &0.72 &1.26 &0.97 &0.73 \\
    \end{tabular*}}{}
     \caption{Rand Index ($RI$) and Dunn Index ($DI$) related to the clustering of randomization methods for various empirical hypergraphs (see Fig. \ref{SI_fig_reshuffling_all_data}). The clusters are determined both by the agglomerative algorithm (``minimum" or ``average") and the dissimilarity metric (HNS, HPD, NS, PD) considered.
    \label{SI_tab:RI_DI_reshuff_datasets}}
\end{table}

\section*{Clustering of empirical hypergraphs}

\begin{figure*}[t!]
    \includegraphics[width=\textwidth]{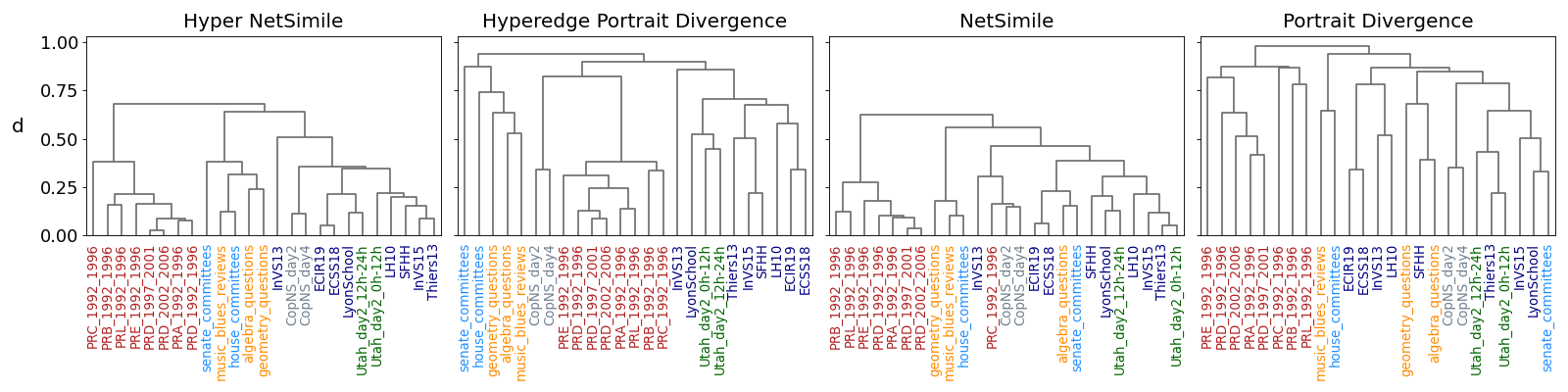}
    \caption{\label{SI_fig_dendrogram_average} 
    \textbf{Clustering of empirical hypergraphs obtained through the \modified{``average"} clustering algorithm.} The dendrograms are obtained with higher-order (HNS, HPD) and pairwise (NS, PD) metrics. The colors of the labels reflect the type of data from which the hypergraphs are extracted: red for co-authorship, yellow for online interactions, light blue for committees membership; all the remaining labels describe face-to-face interactions and the colors reflect the different organizations that collected the data (blue for SocioPatterns, gray for Copenhagen Network Study, and green for Utah’s School-age Population project).}
\end{figure*}

In Fig. \ref{SI_fig_dendrogram_average} we report the dendrograms associated with the clustering of empirical hypergraphs performed via the \modified{``average"} algorithm for the four dissimilarity measures (see Methods in the main text for a full description of the data sets).
As for the \modified{``minimum"} algorithm presented in the main text, the higher-order metrics provide a better classification of the hypergraphs, according to the context they come from.
\modified{In particular, HNS, HPD, and PD separate the group of co-authorship data sets (in red) well from the rest, but only HPD finds that the closest elements are the ones referring to the same journal (PRD) in different years (1992-1996, 1997-2001, 2002-2006).
}
\modified{On the contrary, NS misclassifies one of the co-autorship hypergraphs (PRC\_1992\_1996). Furthermore, the higher-order measures correctly identify the cluster of hypergraphs based on face-to-face interactions, while the pairwise ones tend to mix them with elements of online interactions and committees membership data sets.
}
\modified{The only exception is the case of the hypergraphs built from the Copenhagen Network Study data sets (gray labels) in the dendrogram related to HPD. They are not placed in the same branch as the other face-to-face interactions hypergraphs, probably due to the different way in which the data were collected (see Section``Clustering of real-world hypergraphs" in the main text).
}

\section*{Computational complexity of Hyperedge Portrait Divergence}

\modified{In this section we derive an expression for the computational complexity of Hyperedge Portrait Divergence.
We start by considering an hypergraph $\mathcal{H} = (\mathcal{V}, \mathcal{E})$, with $N = |\mathcal{V}|$ nodes and $E = |\mathcal{E}|$ hyperedges. 
Let us define the auxiliary pairwise graph (also called line graph) $\mathcal{G^*} = (\mathcal{V}^*, \mathcal{E}^*)$, with $N^* = |\mathcal{V}^*|$ nodes and $E^* = |\mathcal{E}^*|$ edges, in which every node represent an hyperedge $e_i$ of $\mathcal{H}$ (i.e., $\mathcal{V}^* = \mathcal{E}$) and two nodes $e_1, e_2$ are connected if the hyperedges they represent in $\mathcal{H}$ overlap with each other, that is, if they share at least one node in $\mathcal{H}$.
To build the hyperedge portrait $\Gamma$ of $\mathcal{H}$, we need to compute the shortest path between all the hyperedges in $\mathcal{E}$ or, equivalently, the shortest paths between all the pairs of nodes in $\mathcal{G}^*$.
Since $\mathcal{G}^*$ is an undirected and unweighted network, the computational complexity of this algorithm is $\mathcal{O}(N^* + E^*)$ \cite{cormen2009}.
Now we want to rephrase this scaling in terms of quantities related to $\mathcal{H}$ instead of $\mathcal{G}^*$.
}

\modified{For the first term we have $N^* = E$ by construction.
To compute the second term, we notice that the degree of a node $e \in \mathcal{V}^*$, denoted by $k_e^*$, is equal to the number of hyperedges overlapping with $e \in \mathcal{E}$ in the original hypergraph representation. This quantity is bounded by the sum of the hyperdegrees of the nodes contained in $e$, minus the size of $e$:
\begin{equation}
    k_e^* \leq \sum_{i \in e} k_i - |e| \ .
    \label{SI_eq_complexity_1}
\end{equation}
The equality in Eq. \eqref{SI_eq_complexity_1} holds only if all the nodes in $e$ are connected to each other only through $e$, that is, if the hyperedges overlapping with $e$ share only a node with $e$.
The above expression allows to compute an upper bound for $E^*$:
\begin{equation}
\begin{split}
    E^* &= \frac{1}{2} \sum_{e \in \mathcal{V}^*} k_e^* \leq 
    \frac{1}{2} \sum_{e \in \mathcal{E}} \bigg( \sum_{i\in e} k_i - |e|  \bigg) =
    \frac{1}{2} \bigg(  \sum_{e \in \mathcal{E}} \sum_{i\in e} k_i - \sum_{e \in \mathcal{E}} |e|  \bigg) = \\ &= 
    \frac{1}{2} \bigg(  \sum_{i \in \mathcal{V}} k_i^2 - \sum_{i \in \mathcal{V}} k_i  \bigg) =
    \frac{1}{2} N \bigg( \langle k^2 \rangle - \langle k \rangle \bigg) \ .
\end{split}
\end{equation}
This implies that in the worst case scenario $E^* = N \big( \langle k^2 \rangle - \langle k \rangle \big) / 2$, and the computational cost of HPD is $\mathcal{O} \big( E + N ( \langle k^2 \rangle - \langle k \rangle) \big)$.
}

\bibliographystyle{naturemag}
\bibliography{ref_supplementary}